\documentclass{svjour2}                    
\smartqed  
\usepackage{graphicx,mathrsfs}
\usepackage{amsfonts,bm,amsmath,amssymb,mathrsfs}
\journalname{Journal of Statistical Physics}

\begin{document}

\title{Dispersion and collapse in stochastic velocity fields on a cylinder}


\author{Antonio~Celani \and Sylvain~Rubenthaler \and Dario~Vincenzi}


\institute{A. Celani \at
              CNRS URA 2171, Institut Pasteur \\
              28 rue du docteur Roux\\
              75724 Paris Cedex 15\\
              France
           \and
           S. Rubenthaler \at
           CNRS UMR 6621, Laboratoire J.A. Dieudonn\'e\\
           Universit\'e de Nice Sophia Antipolis\\
           Parc Valrose\\
           06108 Nice Cedex 2\\
           France              
           \and
           D. Vincenzi \at
           CNRS UMR 6621, Laboratoire J.A. Dieudonn\'e\\
           Universit\'e de Nice Sophia Antipolis\\
           Parc Valrose\\
           06108 Nice Cedex 2\\
           France
}

\date{}

\maketitle

\begin{abstract}
The dynamics of fluid particles on cylindrical manifolds is investigated.
The velocity field is obtained by generalizing the isotropic Kraichnan ensemble, and is therefore
Gaussian and decorrelated in time.
The degree of compressibility is such that when the radius of the cylinder tends to infinity
the fluid particles separate in an explosive way.
Nevertheless, when the radius is finite the transition probability of the two-particle separation
converges to an invariant measure. This behavior is due to the large-scale compressibility
generated by the compactification of one dimension of the space.
\end{abstract}

\section{Introduction}
\label{intro}

Many physical systems display a strong dependence on the space dimensionality, the best known
example being given by phase transitions in equilibrium statistical physics. As for non-equilibrium systems, hydrodynamic turbulence shows a remarkable
dependence on the space dimension as well. In three dimensions, the kinetic energy 
flows from large to small scales in the form of a Kolmogorov--Richardson cascade. Conversely,
in two dimensions, energy is transferred upscale at a constant rate, 
in an inverse cascade process~\cite{K67}. Additionally,
three-dimensional turbulence is characterized by a breakdown of scaling invariance and small-scale
intermittency \cite{F95}, whereas the inverse cascade is apparently self-similar \cite{BCV00} and even shows some intriguing
signatures of conformal invariance \cite{BBCF06}. 

These observations have spurred the search of a critical dimension
between $d=2$ and $d=3$ in the hope that it could provide  a starting point for an analytical attack
of three- or two-dimensional turbulence, or both. This approach has been mainly applied to
simplified models of turbulence, such as EDQNM \cite{FF78}, the shell model
\cite{GJY02}, or a model obtained by generalizing the form of the 
two-dimensional stream function~\cite{LPP02}.
In those studies, the spatial dimension has been most conveniently 
reduced to a formal parameter that could take arbitrary values.
The approach undertaken in the present paper differs from previous work
at least in two important aspects. First, we shall consider a geometrical, rather than formal, way of 
looking in between dimensions. Namely, we shall study the dynamics of fluid particles 
on cylindrical manifolds where the compactified dimension can be 
collapsed or inflated at will so as to connect continuously the two extreme cases. Second, we 
shall focus on a system that is fully under analytical control, that is 
the Kraichnan ensemble of velocities rather than Navier--Stokes turbulence. 

To study the turbulent transport of a passive scalar, Kraichnan introduced
a Gaussian ensemble of decorrelated-in-time velocity fields~\cite{K68}.
A compressible generalization of the Kraichnan ensemble in the $d$-dimen\-sional Euclidean space
has been investigated by Gaw\c{e}dzki and
Vergassola under the assumption of statistical isotropy~\cite{GV00}.%
\footnote{The smooth limit of this model had been previously considered in ref.~\cite{CKV98}.} 
In this model, 
the dynamics of fluid particles depends on three physical
quantities: the space dimension, the degree of compressibility,
and the (spatial) H\"older exponent of the velocity.
The H\"older exponent~$\xi/2$ is greater than zero and less than one.
This property mimics the behavior
of a turbulent velocity field, whose realizations are typically non-Lipschitz
in the limit of infinite Reynolds number.
For any given~$d<4$ and~$0<\xi<2$, 
Gaw\c{e}dzki and Vergassola have identified
a critical degree of compressibility separating
two different phases of the Lagrangian dynamics.
Below the critical value (incompressible or weakly compressible velocity 
fields), fluid particles separate superdiffusively.
The probability distribution of fluid-particle separations does not have
a stationary limit in this regime. 
Above the threshold (strongly compressible fields), Lagrangian
trajectories tend to collapse to zero distance, and
the distribution of the separations degenerates into
a Dirac delta function.
For~$d\geq 4$, the former regime is the only possible one and the phase transition does not occur.%
\footnote{For~$d=4$, 
the collapsed phase can exist only for smooth velocity fields ($\xi=2$).
}
The above results have been subsequently elaborated by Le Jan and Raimond
in the context of non-Lipschitz stochastic differential equations~\cite{LJR02,LJR04}.

Here we consider a generalization of the Kraichnan ensemble on a cylindrical surface.
A $d$-dimensional cylindrical surface can be constructed by taking~$\mathbb{R}^d$ 
and compactifying $d-d'$ dimensions. The radius of the cylinder
is the size of the compactified dimensions. When the radius tends to 
infinity we recover~$\mathbb{R}^d$; 
when it tends to zero we obtain~$\mathbb{R}^{d'}$.
Thus, varying the radius of
the cylinder produces a smooth transformation 
from dimension~$d$ to dimension~$d'$. 

We define a zero-mean Gaussian velocity field on a cylindrical surface
by imposing the form of its covariance. We require that the covariance of the field
tends to the one of the isotropic $d$-dimensional Kraichnan ensemble
as the radius of the cylinder tends to infinity 
and to the one of the isotropic $d'$-dimensional Kraichnan ensemble 
as the radius vanishes. 
The degree of compressibility is such that the velocity
is weakly compressible in the limit of infinite radius and
strongly compressible in the opposite limit. 
It is therefore possible to gradually move from one regime to the other by varying the radius of the
cylinder. 

As we shall see, if in the limit of infinite radius
the H\"older exponent  is equal to~$\xi$, 
then in the limit of vanishing radius it is equal to~$\xi+(d-d')$.
Hence, if attention is restricted
to non-smooth velocities, the model under consideration
is meaningful only when a single dimension is compactified ($d'=d-1$).
For the sake of simplicity, we shall conduct the analysis in two dimensions, where the two extreme cases
are the two-dimensional plane and the straight line.
We shall show that
the dynamics of fluid particles results from two opposite effects.
At small separations, Lagrangian trajectories exhibit a superdiffusive dynamics owing to the
weakly compressible nature of the small-scale velocity.
At large separations, fluid particles experience the trapping effect of a strongly compressible
field.  Consequently,
the probability distribution of the two-particle separation tends to an invariant measure.
This behavior is to be contrasted with the one observed in the two-dimensional isotropic case with
the same H\"older exponent and degree of compressibility.

In the present context,
the separation vector between two fluid particles is a stochastic process
solving an It\^o stochastic differential equation with non-Lipschitz diffusion
coefficient. To guarantee the existence and the uniqueness in law of the solution,
we shall add pure diffusion to the velocity field.
By considering an appropriate Lyapunov function, we shall demonstrate that there exists an invariant
measure for the fluid-particle separation. The invariant measure is unique, ergodic, and 
non-degenerate
as a consequence of the irreducibility and the strong Feller property of the process.

The paper is divided as follows.
Section~\ref{sec:model} describes a generalization of the Kraichnan model on a $d$-dimensional 
cylindrical surface. The two-dimensional case is studied in detail in section~\ref{sec:2D}.
Sections~\ref{sec:4}
and~\ref{sec:5} contain the results on the fluid-particle separation and its
invariant measure.
The limit of vanishing diffusivity, the effect of a viscous regularization
of the velocity field, and the role of the Prandtl number are discussed in section~\ref{sec:conclusions}.

\section{Kraichnan model on a $d$-dimensional cylindrical surface}
\label{sec:model}

We consider the dynamics of fluid particles in a turbulent flow
on a $d$-dimen\-sional cylindrical surface~$\mathcal{S}$.
The velocity field is a family of white noises taking their
values in the space of vector fields on~$\mathcal{S}$.
Specifically, $\bm v(t,\bm x)$ is a Gaussian stochastic process with
zero mean and covariance
\begin{equation}
\label{eq:covariance}
\mathbb{E}(v_\alpha(t,\bm x)v_\beta(s,\bm y))=
D_{\alpha\beta}(\bm x-\bm y)\delta(t-s),
\end{equation}
where~$\bm x,\bm y\in\mathbb{R}^{d'}\times[-\pi L,\pi L)^{d-d'}\subset 
\mathbb{R}^d$ ($d>d'$) and~$L$ is the radius of the cylinder.  
The velocity is by definition
statistically homogeneous in space, stationary in time, and
invariant under time reversal.
Moreover, we assume periodicity in the~$d-d'$ ``radial'' coordinates.

It is convenient to write
the spatial covariance~$D_{\alpha\beta}(\bm r)$ 
in terms of its Fourier-space representation: 
\[
D_{\alpha\beta}(\bm r)=\frac{1}{(2\pi)^{d'}(2\pi L)^{d-d'}}\sum_{\bm k''\in \frac{1}{L}\mathbb{Z}^{d-d'}} e^{i\bm k''\cdot\bm r''}\int_{\mathbb{R}^{d'}} d\bm k'\,
e^{i\bm k'\cdot\bm r'}\, F_{\alpha\beta}(\bm k)
\]
with~$\bm k=(\bm k',\bm k'')\in \mathbb{R}^{d'}\times \frac{1}{L}
\mathbb{Z}^{d-d'}$, 
$\bm r=(\bm r',\bm r'')\in\mathbb{R}^{d'}\times[-\pi L,\pi L)^{d-d'}$,
and~$\alpha,\beta=1,\dots,d$.
The presence of a series in the~$\bm k''$-coordinates accounts for the
periodicity of the velocity field in the~$\bm r''$-coordinates.

We adopt the following form for the spectral tensor:
\begin{equation}
\label{eq:spectral-tensor}
F_{\alpha\beta}(\bm k)=
\frac{A_{\alpha\beta}(\bm k;\wp\big)}%
{(\Vert\bm k\Vert^2+\ell^{-2})^{\frac{d+\xi}{2}}}
\end{equation}
with $\ell\in\mathbb{R}_+$, $\xi\in[0,2]$, $\wp\in [0,1]$, and
\[
A_{\alpha\beta}(\bm k;\wp)=
(1-\wp)\delta_{\alpha\beta}
+(\wp d-1)\,\frac{k_\alpha k_\beta}{\Vert \bm k\Vert^2}. 
\]
As we shall see in the latter part of this section,
$F_{\alpha\beta}(\bm k)$ has been chosen in such a way 
that, in the limits~$L\to 0$ and~$L\to\infty$, $D_{\alpha\beta}(\bm r)$
tends to the covariance of an isotropic random
field.\footnote{The spectral tensor could in principle be 
multiplied by a positive coefficient
determining the intensity of the velocity fluctuations. For the sake of
simplicity, we set that coefficient to one.}

The spectral tensor is real, symmetric ($F_{\alpha\beta}(\bm k)=
F_{\beta\alpha}(\bm k)$) and non-negative definite
$\forall\, \bm k\in \mathbb{R}^{d'}\times \frac{1}{L}\mathbb{Z}^{d-d'}$, i.e.,
\[
\sum_{1\leq\alpha,\beta\leq d}F_{\alpha\beta}(\bm k)u_\alpha u_\beta \geq 0
\qquad \forall\, (u_1,\dots,u_d) \in\mathbb{R}^d,
\]
as can be checked using the Cauchy--Schwartz inequality. 
These properties guarantee that $D_{\alpha\beta}(\bm r)$ is the 
spatial covariance of a homogeneous random field (e.g., ref.~\cite{MY75}, p.~20).
Moreover, $F_{\alpha\beta}(\bm k)$ is an even function 
of~$\bm k$, and therefore the velocity is
statistically invariant under parity:
$D_{\alpha\beta}(-\bm r)=D_{\alpha\beta}(\bm r)$. 

As a consequence of statistical homogeneity and parity invariance, 
the covariance of velocity differences can be expressed in terms 
of~$D_{\alpha\beta}(\bm r)$:
\begin{equation}
\label{eq:structure}
\mathbb{E}([v_\alpha(t,\bm x+\bm r)-v_\alpha(t,\bm x)][v_\beta(s,\bm x+\bm r)-
v_\beta(s,\bm x)])=
2d_{\alpha\beta}(\bm r)\delta(t-s)
\end{equation}
with~$d_{\alpha\beta}(\bm r)=D_{\alpha\beta}(0)-D_{\alpha\beta}(\bm r)$~\cite{MY75}. 

The meaning of the parameters~$\wp$, $\ell$, and~$\xi$
may be understood
by considering the limit of~$D_{\alpha\beta}(\bm r)$
for~$L\to \infty$ and for~$L\to 0$.

The limit~$L\to\infty$ (and~$1/L\to d\bm k''$) yields: 
\begin{multline}
\label{eq:L-to-inf}
\lim_{L\to\infty}D_{\alpha\beta}(\bm r)=
\dfrac{1}{(2\pi)^d}\int_{\mathbb{R}^{d-d'}}\!d\bm k''
\int_{\mathbb{R}^{d'}}d\bm k'\,
\frac{e^{i\bm k'\cdot\bm r'+i\bm k''\cdot\bm r''}A_{\alpha\beta}\left(\bm k;\wp\right)}%
{\left(\Vert\bm k'\Vert^2+\Vert\bm k''\Vert^2 +
\ell^{-2}\right)^{\frac{d+\xi}{2}}}\
\\[0.2cm]
=\dfrac{1}{(2\pi)^d}\int_{\mathbb{R}^d}d\bm k\,
\frac{e^{i\bm k\cdot\bm r}A_{\alpha\beta}(\bm k;\wp)}%
{\left(\Vert\bm k\Vert^2+\ell^{-2}\right)^{\frac{d+\xi}{2}}}\, .
\end{multline}
In this limit, $D_{\alpha\beta}(\bm r)$
tends  to the spatial covariance of a 
$d$-dimensional isotropic field 
with correlation length~$\ell$ and degree of 
compressibility~$\wp$~\cite{GV00,FGV01}. 
The parameter~$\xi/2$ represents the inertial-range H\"older exponent 
of the velocity: 
$\sum_{\alpha=1}^d d_{\alpha\alpha}(\bm r)=O(\Vert\bm r\Vert^\xi)$
as~$\Vert\bm r/\ell\Vert\to 0$.
For~$\xi=0$ the velocity field is purely diffusive;
for $\xi=2$ it is spatially smooth, and its spatial regularity
decreases with decreasing~$\xi$.
In particular, the Kolmogorov scaling 
is obtained for~$\xi=4/3$, for
the time integral of~\eqref{eq:structure}
must be proportional to~$\Vert\bm r\Vert^{4/3}$
in Kolmogorov's phenomenology~\cite{F95}.%
\footnote{The same conclusion can be reached rigorously by defining
the Kraichnan ensemble as the limit of an Ornstein--Uhlenbeck process 
for vanishing correlation time~\cite{F04}.}

It is worth noting that for a finite~$L$ equation~\eqref{eq:L-to-inf}
describes the velocity covariance
at space separations much smaller than~$L$.

In the second limit, $L\to 0$, we obtain%
\footnote{This can be shown by multiplying~$D_{\alpha\beta}(\bm r)$ 
by a function~$f(\bm r''$),
integrating over~$\bm r''\in [-\pi L,\pi L)$, taking the limit~$L\to 0$, and noting that only 
the term corresponding to $\bm k''=0$ has a non-zero limit equal to
$$
f(0)\,\dfrac{\mathcal{K}}{(2\pi)^{d'}}
\int_{\mathbb{R}^{d'}}d\bm k'\,
\frac{e^{i\bm k'\cdot\bm r'}A_{\alpha\beta}(\bm k';\wp')}%
{\left(\Vert\bm k'\Vert^2+\ell^{-2}\right)^{\frac{d'+\xi'}{2}}}\, .
$$}
$$
\lim_{L\to 0}D_{\alpha\beta}(\bm r)=
\delta(\bm r'')\,\dfrac{\mathcal{K}}{(2\pi)^{d'}}
\int_{\mathbb{R}^{d'}}d\bm k'\,
\frac{e^{i\bm k'\cdot\bm r'}A_{\alpha\beta}(\bm k';\wp')}%
{\left(\Vert\bm k'\Vert^2+\ell^{-2}\right)^{\frac{d'+\xi'}{2}}}
$$
with $\xi'=\xi+(d-d')$ and
\begin{align*}
&\wp'=\dfrac{\wp(d-1)}{\wp(d-d')+d'-1}, & 
\mathcal{K}&=1+\,\dfrac{\wp(d-d')}{d'-1} & \text{if $d'>1$},&
\\[0.3cm]
&A_{11}(k';\wp')=1, & \mathcal{K}&=\wp (d-1) & \text{if $d'=1$ and $\wp> 0$}&.
\end{align*} 
We thus recover the
covariance of a $d'$-dimensional isotropic velocity field
with H\"older exponent~$\xi'/2$, correlation length~$\ell$,
and degree of compressibility~$\wp'$~\cite{GV00,FGV01}.

The exponents~$\xi$ and $\xi'$ must satisfy the inequalities $0\le\xi\le 2$ and $0\le\xi'\le 2$.
Therefore, the limit~$L\to 0$ makes sense only in two cases:
\begin{itemize}
\item[\textit{a})]
 $d'=d-1$, $\xi\in[0,1]$, and $\xi'=\xi+1 \in [1,2]$;
\item[\textit{b})]
 $d'=d-2$, $\xi=0$, and $\xi'=2$.
\end{itemize}
We are interested in the situation where the
fluid particles disperse
when~$L\to\infty$ ($d$-dimensional isotropic flow) and 
collapse when~$L\to 0$
($d'$-dimensional isotropic flow). Moreover, we focus on spatially rough
velocity fields leaving aside the cases~$\xi=0$ and~$\xi'=2$.
This situation can be realized only in case~\textit{a}), 
for~$\xi\in(0,1)$, and under the conditions~\cite{GV00}:
$$
\wp < \dfrac{d}{\xi^2} \qquad \text{and} \qquad \wp'\ge\dfrac{d'}{{\xi'}^2}.
$$
The first inequality is actually satisfied for all~$d$ and~$\xi$ given 
that~$\xi\in(0,1)$ and~$\wp\in[0,1]$.
The second inequality can be rewritten in terms of~$\wp$ as follows:
\begin{equation}
\label{eq:compressibility}
\wp \ge \dfrac{d-2}{\xi(\xi+2)}. 
\end{equation}
The restriction~$0\le\wp\le 1$ and inequality~\eqref{eq:compressibility}
imply the additional condition~$d<5$.

In the remainder of the paper, we shall investigate the 
statistics of fluid-particle separations on a 
two-dimensional cylindrical surface ($d=2$).

\section{Two-dimensional cylindrical surface}
\label{sec:2D}
For $d=2$, case \textit{a}) is the only realizable one,
corresponding to~$d'=1$.
Condition~\eqref{eq:compressibility} reduces
to~$\wp\ge 0$ independently of~$\xi$.

The spatial covariance of the velocity field takes the form
\begin{equation}
\label{eq:2d}
D_{\alpha\beta}(\bm r)=\frac{1}{4\pi^2 L}
\sum_{j=-\infty}^{\infty} e^{i\frac{j}{L}r_2}\int_{\mathbb{R}} dk_1\,
\frac{e^{ik_1 r_1}A_{\alpha\beta}((k_1,\frac{j}{L});\wp)}%
{\big[k_1^2+\big(\frac{j}{L}\big)^2+\frac{1}{\ell^{2}}\big]^{\frac{2+\xi}{2}}}
\end{equation}
with~$\bm r=(r_1,r_2)\in \Omega=\mathbb{R}\times[-\pi L,\pi L)$.
In eq.~\eqref{eq:2d}
we have written~$\bm k=(k_1,k_2)$ 
with~$k_2=j/L$ and~$j\in\mathbb{Z}$
to make the dependence on~$L$ explicit. We shall keep this notation in the
remainder of the paper.

We now restrict attention to 
space separations much smaller than~$\ell$. Formally, this is 
equivalent to considering the limit~$\ell\to\infty$.
The spatial variance of the velocity field, $D_{\alpha\beta}(\bm 0)$, 
diverges as~$\ell$ tends to infinity (appendix~\ref{app:A}); 
this behavior reflects the divergence of the
average kinetic energy of the fluid.
Nevertheless, $d_{\alpha\beta}(\bm r)$ has a finite limit for all~$\bm r$, 
and the statistics of velocity differences remains well defined.

The limit of~$d_{\alpha\beta}(\bm r)$ for~$\ell\to\infty$
can be computed explicitly (appendix~\ref{app:A}).
The correlation of the axial component is written:
\begin{multline}
\label{eq:d11}
\lim_{\ell\to\infty}
d_{11}({\bm r})= 
\frac{\wp\left|\Gamma\left(-\frac{1+\xi}{2}\right)\right|}
{2^{3+\xi}\pi^{3/2}\Gamma\left(1+\frac{\xi}{2}\right)L}\,\vert r_1\vert^{1+\xi}
\\[2mm]
+ \frac{L^{\xi}}{2\pi^{3/2}\Gamma\left(2+\frac{\xi}{2}\right)}
\displaystyle\sum_{j=1}^{\infty} j^{-1-\xi} \left\{
\frac{1+(1-\wp)\xi}{2}\,\Gamma\left(\frac{1+\xi}{2}\right)\right.
\\[0.2cm]
\displaystyle - 2 \cos\left(\frac{jr_2}{L}\right) \left[\wp\left(1+\frac{\xi}{2}\right) 
\left(\frac{j\vert r_1\vert}{2L}\right)^{\frac{1+\xi}{2}} 
K_{\frac{1+\xi}{2}}\left(\frac{j\vert r_1\vert}{L}\right)  
\right.
\\
\left.\left.
+(1-2\wp) \left(\frac{j\vert r_1\vert}{2L}\right)^{\frac{3+\xi}{2}}
K_{\frac{3+\xi}{2}}\left(\frac{j\vert r_1\vert}{L}\right)   
\right]
\right\},
\end{multline}
where~$K_\nu(z)$ denotes the
modified Bessel function
of the second kind of order~$\nu$ and argument~$z$.
The correlation of the radial component has the form:
\begin{multline}
\label{eq:d22}
\lim_{\ell\to\infty}
d_{22}({\bm r})= 
\frac{(1-\wp)\left|\Gamma\left(-\frac{1+\xi}{2}\right)\right|}
{2^{3+\xi}\pi^{3/2}\Gamma\left(1+\frac{\xi}{2}\right)L}\,\vert r_1\vert^{1+\xi}
\\[2mm]
+\frac{L^{\xi}}{2\pi^{3/2}\Gamma\left(2+\frac{\xi}{2}\right)}
\displaystyle\sum_{j=1}^{\infty} j^{-1-\xi} 
\left\{\frac{1+\wp\xi}{2}\,\Gamma\left(\frac{1+\xi}{2}\right)\right.
\\[0.2cm] 
\displaystyle - 2 \cos\left(\frac{jr_2}{L}\right) \left[(1-\wp)\left(1+\frac{\xi}{2}\right) 
\left(\frac{j\vert r_1\vert}{2L}\right)^{\frac{1+\xi}{2}} 
K_{\frac{1+\xi}{2}}\left(\frac{j\vert r_1\vert}{L}\right)  \right.
\\
\left.\left.
+(2\wp-1) \left(\frac{j\vert r_1\vert}{2L}\right)^{\frac{3+\xi}{2}}
K_{\frac{3+\xi}{2}}\left(\frac{j\vert r_1\vert}{L}\right)   
\right]\right\}.
\end{multline}
Finally, the mixed correlations can be written as follows:
\begin{multline}
\label{eq:d12}
\lim_{\ell\to\infty}
d_{12}({\bm r})= \lim_{\ell\to\infty}d_{21}(\bm r)
\\ \displaystyle
=\frac{(2\wp-1)L^{\xi} }{2\pi^{3/2}\Gamma\left(2+\frac{\xi}{2}\right) }
\sum_{j=1}^{\infty} j^{-1-\xi} \left(\frac{j r_1}{L}\right)  
\sin\left( \frac{jr_2}{L}\right)
\left(\frac{j\vert r_1\vert}{2L}\right)^{\frac{1+\xi}{2}}
K_{\frac{1+\xi}{2}}\left(\frac{j\vert r_1\vert}{L}\right).
\end{multline}
The limit~$\ell\to\infty$ will be hereafter understood.

\subsection{Large-scale form of the covariance of velocity differences}
\label{sec:large-scale}

To understand the nature of the random velocity field, 
it is useful to consider the covariance of velocity differences
at space separations much greater than the radius of the cylinder.
 
The series in eqs.~\eqref{eq:d11}-\eqref{eq:d12} converge
uniformly (appendix~\ref{app:A}). 
For~$\vert r_1\vert/L\to\infty$,
it is therefore possible to deduce
the asymptotic expansion of~$d_{\alpha\beta}(\bm r)$ 
from the limiting behavior of the single terms of the series.
The asymptotic expansion of~$K_\nu(z)$  for~$z\to\infty$ is
(e.g., ref.~\cite{E53}, formula~II~7.13(7))
\[
K_\nu(z)\sim\sqrt{\frac{\pi}{2z}}\, e^{-z} 
\qquad (\vert\arg z\vert<3\pi/2).
\]
Thus, the $r_2$-dependent 
contributions to~$d_{\alpha\beta}(\bm r)$ decay exponentially fast
with increasing space separation.
The remaining contributions give 
\begin{equation}
\label{eq:d11-asymptotic}
d_{11}(\bm r)\sim \mathfrak{D}_1\,\vert r_1\vert^{1+\xi}+\varkappa_1
\qquad \text{as $\dfrac{\vert r_1\vert}{L}\to\infty$}
\end{equation}
with
\[
\mathfrak{D}_1=\frac{\wp\left|\Gamma\left(-\frac{1+\xi}{2}\right)\right|}%
{2^{3+\xi}\pi^{3/2}\Gamma\left(1+\frac{\xi}{2}\right)L}
\]
and
\[
\varkappa_1=\frac{[1+(1-\wp)\xi]L^{\xi}\Gamma\left(\frac{1+\xi}{2}\right)}%
{2\pi^{3/2}(2+\xi)\Gamma\left(1+\frac{\xi}{2}\right)}\,\zeta(1+\xi).
\]
In the latter equation
$$
\zeta(s)=\displaystyle\sum_{j=1}^{\infty} \frac{1}{j^{s}} \qquad (s>1)
$$
is the Riemann Zeta function. Likewise, we have 
\begin{equation}
\label{eq:d22-asymptotic}
d_{22}(\bm r) \sim \mathfrak{D}_2\,\vert r_1\vert^{1+\xi}+\varkappa_2
\qquad \text{as $\dfrac{\vert r_1\vert}{L}\to\infty$}
\end{equation}
with
\[
\mathfrak{D}_2=\dfrac{1-\wp}{\wp}\,\mathfrak{D}_1, \qquad \varkappa_2=\dfrac{1+\wp\xi}{1+(1-\wp)\xi}\,\varkappa_1.
\]
Finally, the off-diagonal terms vanish at large space separations:
\begin{equation}
\label{eq:d12-asymptotic}
\lim_{\vert r_1\vert/L\to\infty} d_{12}({\bm r})=
\lim_{\vert r_1\vert/L\to\infty}d_{21}(\bm r) =0.
\end{equation}
The above asymptotic expressions show that, at separations much greater than
the radius of the cylinder, the velocity difference may be regarded as
the superposition of
two independent one-dimensional random fields. 
One field is directed along the axial direction; the other one is
directed along the radial direction. Both the fields depend only on~$r_1$.
In particular, the axial field is a  one-dimensional Kraichnan
velocity field with H\"older exponent~$1+\xi$ like the one
considered in ref.~\cite{VM97}.

At large separations, the
small-scale dynamics manifests itself through
an effective diffusivity represented 
by the constants~$\varkappa_1$ and~$\varkappa_2$.

\section{Fluid-particle dynamics}
\label{sec:4}

In the present context,
 the separation between two fluid particles
can be regarded as a stochastic process on~$\Omega$ with diffusion
coefficient~$d_{\alpha\beta}(\bm r)$ (and drift coefficient equal to zero). To ensure the (weak) existence and uniqueness
of the trajectories of the process, we add diffusion to the velocity field
and replace~$d_{\alpha\beta}(\bm r)$ by
$$
d_{\alpha\beta}^\kappa(\bm r):=d_{\alpha\beta}(\bm r)+
2\kappa\delta_{\alpha\beta}, 
\qquad \kappa>0.
$$
The additional term can model the action of  molecular diffusion on fluid particles as, e.g., in ref.~\cite{GV00}.
The constant~$\kappa$ will be referred to as diffusivity.

The separation vector between two fluid particles will be denoted by~$\bm R$.
According to the above remark, $\bm R$
satisfies the It\^o stochastic differential equation%
\footnote{If~$\bm X$ and~$\bm Y$ denote the positions of two fluid particles, the separation vector
between the two particles is defined as~$\bm R:=\bm Y-\bm X$. The common physical notation
for the evolution equation for~$\bm R$ would be
$$
\dfrac{d\bm R}{dt}=\delta_{\bm R}\bm v(t)+2\sqrt{\kappa}\,\bm\xi(t) 
$$
where~$\bm\xi$ is white noise and 
the statistics of~$\delta_{\bm R}\bm v(t):=\bm v(t,\bm Y(t))-\bm v(t,\bm X(t))$ is defined by 
eq.~\eqref{eq:structure}}
\begin{equation}
\label{eq:weak}
d\bm R(t)=
\sqrt{2}\,\sigma(\bm R(t))d\bm B(t), \qquad \bm R(0)=\bm r \in\Omega,
\end{equation}
where~$\bm B$ is Brownian motion on~$\Omega$ and~$\sigma$
is defined through the Cholesky decomposition of the matrix~$d^\kappa$:
$$\sigma\sigma^\mathrm{T}=d^\kappa$$ 
with 
\[
\sigma(\boldsymbol{r})=
\begin{pmatrix}
\sqrt{d_{11}^\kappa(\boldsymbol{r})} && 0
\\[2mm]
\dfrac{d_{12}^\kappa(\bm r)}{\sqrt{d_{11}^\kappa(\bm r)}}
&& 
\sqrt{d_{22}^\kappa(\bm r)-\dfrac{[d_{12}^\kappa(\bm r)]^2}{d_{11}^\kappa(\bm r)}}
\end{pmatrix}, \qquad \bm r\in\Omega.
\]
Although the diffusion matrix~$d^\kappa$
is nor Lipschitz continuous nor bounded, the existence and uniqueness of the
solution of eq.~\eqref{eq:weak} can be proved using 
Stroock's and Varadhan's theory of martingale problems~\cite{SV79}.
To directly exploit this theory,
we shall first consider the periodic extension of 
eq.~\eqref{eq:weak}
on~$\mathbb{R}^2$, and then project the resulting process on~$\Omega$.
We therefore introduce the projection~$\mathfrak{p}:\mathbb{R}^2\to\Omega$ with
\[
\mathfrak{p}(\bm r)=\left(r_1,-\pi L+r_2-2\pi L 
\left\lfloor\frac{r_2}{2\pi L}\right\rfloor\right),
\]
and define~$\tilde{\sigma}:=\sigma\circ\mathfrak{p}$.
Likewise we denote~$\tilde{d}^\kappa:=d^\kappa\circ\mathfrak{p}=
\tilde{\sigma}\tilde{\sigma}^\mathrm{T}$.

Before proceeding further, it is convenient to define some notation.
The spaces of bounded measurable and bounded continuous functions 
on~$\Omega$ will be
denoted by~$\mathscr{B}_b(\Omega)$ and~$\mathscr{C}_b(\Omega)$, respectively.
The set~$\mathscr{C}^2(\Omega)$ will be the space
of functions having two continuous derivatives.
Analogous definitions will apply to functions defined on~$\mathbb{R}^2$.
\begin{proposition}
The It\^o stochastic differential equation on~$\mathbb{R}^2$:
\begin{equation}
\label{eq:weak-r2}
d\widetilde{\bm R}(t)=
\sqrt{2}\,\tilde{\sigma}
\big(\widetilde{\bm R}(t)\big)d\widetilde{\bm B}(t), 
\qquad \widetilde{\bm R}(0)=\bm r\in\mathbb{R}^2,
\end{equation} 
where~$\widetilde{\bm B}$ 
is the standard Brownian motion on~$\mathbb{R}^2$, 
has a unique (in law) weak solution.
In particular, the solution is a continuous Markov process.

For~$\bm r\in\mathbb{R}^2$ and~$U\subseteq\mathbb{R}^2$ measurable,
let
$$
\widetilde{P}(0,\bm r;s,U):=\mathbb{P}
\left(\widetilde{\bm R}(s)\in U
\text{ $\mathrm{if}$ $\widetilde{\bm R}(0)=\bm r$}\right)
$$ 
be the transition probability distribution of~$\widetilde{\bm R}$, and
let $\big(\widetilde{T}_t\big)_{t\ge 0}$ be the associate transition semigroup:
\[
\widetilde{T}_t f(\bm r):=\int_{\mathbb{R}^2}
f(\bm\rho)\widetilde{P}(0,\bm r;t,d\bm\rho)
\]
with~$f\in\mathscr{B}_b(\mathbb{R}^2)$. 
Then, the semigroup~$\big(\widetilde{T}_t\big)_{t\ge 0}$
has the strong Feller property, 
i.e.~$\widetilde{T}_t(\mathscr{B}_b(\mathbb{R}^2))\subset 
\mathscr{C}_b(\mathbb{R}^2)$ for all $t>0$.
\end{proposition}
\begin{proof}
The diffusion matrix has the following properties:
\begin{enumerate}
\item $\tilde{d}^\kappa$ is continuous;
\item $\tilde{d}^\kappa(\bm r)$ is symmetric and 
strictly positive definite for all~$\bm r\in\mathbb{R}^2$.
The diffusivity~$\kappa$ is indeed assumed to be strictly positive, and
the spatial covariance of velocity differences
must be symmetric and uniformly non-negative definite for 
all~$\bm r$ (ref.~\cite{MY75}, p.~97), 
i.e.,
\[
\sum_{1\leq \alpha,\beta\leq 2} d_{\alpha\beta}(\bm r)u_\alpha
u_\beta\geq 0 \qquad \forall\, (u_1,u_2) \text{ and } \bm r\in\mathbb{R}^2;
\]
\item there exists a positive constant~$C_1$ such that for all~$\alpha$, $\beta$, 
and~$\bm r$
\begin{equation}
\label{eq:d-asympt}
\vert \tilde{d}^\kappa_{\alpha\beta}(\bm r)\vert\leq C_1(1+\Vert\bm r\Vert^2).
\end{equation}
This property is a consequence of the asymptotic 
behaviors~\eqref{eq:d11-asymptotic} to~\eqref{eq:d12-asymptotic} and of the
the fact that~$\tilde{d}^\kappa(\bm r)$ is bounded
at the origin and continuous on~$\mathbb{R}^2$. 
\end{enumerate}
Under the above conditions,
Stroock's and Varadhan's uniqueness theorem apply to the
martingale problem for~$\tilde{d}^\kappa$~\cite{SV79}. 
Then, the proposition follows from the equivalence
between the 
well-posedness of martingale problems and the existence and uniqueness in law
of weak solutions of 
stochastic differential equations (ref.~\cite{RW79},
pp.~159 and~170).
\qed
\end{proof}
The process~$\bm R$ can be regarded as the
projection of~$\widetilde{\bm R}$ on~$\Omega$: $\bm R(t)=\mathfrak{p}\big(\widetilde{\bm R}(t)\big)$.
The properties of~$\bm R$ can then be deduced from those of~$\widetilde{\bm R}$.
\begin{corollary}
\label{cor:R}
Equation~\eqref{eq:weak} has a unique (in law) weak solution.
The transition semigroup~$(T_t)_{t\ge 0}$ associated 
with~$\bm R$ has the strong Feller property: 
$T_t(\mathscr{B}_b(\Omega))\subset \mathscr{C}_b(\Omega)$
for all~$t>0$.
\end{corollary}
\begin{proof}
Given an initial condition~$\bm r\in\Omega$, a weak solution of eq.~\eqref{eq:weak} can
be constructed by taking a solution of eq.~\eqref{eq:weak-r2} with the same initial condition
and projecting it on~$\Omega$.

The key observation to prove 
uniqueness in law is that
any solution of eq.~\eqref{eq:weak} on~$\Omega$
can be uniquely mapped into a continuous solution of eq.~\eqref{eq:weak-r2}
on~$\mathbb{R}^2$. Then, uniqueness in law in~$\mathbb{R}^2$ guarantees that
also the solution on~$\Omega$ is unique in law.

Finally, $\big(\widetilde{T}_t\big)_{t\geq 0}$ has the strong Feller property and 
the projection~$\mathfrak{p}$ is
locally invertible (with continuous inverse). Hence, $(T_t)_{t\ge 0}$ has the strong
Feller property.
\qed
\end{proof}

\section{Invariant measure of fluid-particle separations}
\label{sec:5}

An invariant measure for~$\bm R$ can be constructed by adapting
to the case under examination the procedure described in ref.~\cite{P01}.
Clearly, an invariant measure may exist only if the trajectories of the
stochastic process do not ``escape to infinity''.
To control the behavior of the first component of~$\bm R$, which is not bounded,
we therefore introduce the Lyapunov function~$V:\Omega\to\mathbb{R}_+$:
\[
V(\bm r)=
\begin{cases}
\dfrac{h(h+1)c^4+2(1-h^2)c^2 r_1^2-h(1-h)r_1^4}{4(1-h)
c^{2(h+1)}} & \text{if $\vert r_1\vert\le c$},
\\[4mm]
\dfrac{1}{2(1-h)} \, \vert r_1\vert^{2(1-h)}
& \text{if $\vert r_1\vert> c$},
\end{cases}
\]
where~$c>0$ and~$1>h>0$. The function~$V$ is twice 
continuously differentiable
and has the asymptotic behavior needed for the proof.
\begin{lemma} 
\label{lemma-V} 
If~$(1+\xi)/2>h>1/2$ and~$\mathcal{A}$ denotes the infinitesimal generator 
of~$\left({T}_t\right)_{t\ge 0}$,
then the Lyapunov function has the following properties 
for all~$\xi\in(0,1)$:
\begin{enumerate}
\item $\lim_{\Vert\bm r\Vert\to\infty}\mathcal{A}V(\bm r)=-\infty$;
\item there exists~$m\in\mathbb{R}$ such that~$\mathcal{A}V(\bm r)\le m$
for all~$\bm r\in\Omega$; 
\item
$T_t V(\bm r)=V(\bm r)+\displaystyle \int_0^t T_s\mathcal{A}V(\bm r)ds$.
\end{enumerate}
\end{lemma}
\begin{proof}
For~$f\in\mathscr{C}^2({\Omega})$, 
the infinitesimal generator of~$\left({T}_t\right)_{t\ge 0}$ has the form
\[
\mathcal{A}f(\bm r)=
\mathrm{tr}
[\sigma(\bm r)\sigma^\mathrm{T}(\bm r)D^2f(\bm r)],
\]
where~$D^2 f$ denotes the Hessian of the function~$f$.
The action of the generator~$\mathcal{A}$ on~$V(\bm r)$ is written:
\[
\mathcal{A} V(\bm r)=
\sigma_{11}^2(\bm r)\,\dfrac{\partial^2 V}{\partial r_1^2}.
\]
From eq.~\eqref{eq:d11-asymptotic} we obtain
\[
\mathcal{A}V(\bm r)\sim 
-(2h-1)\mathfrak{D}_1\vert r_1\vert^{1+\xi-2h}
\qquad \text{as $\Vert \bm r\Vert\to\infty$}.
\]
If~$(1+\xi)/2>h>1/2$, we have
$\lim_{\Vert\bm r\Vert\to\infty}\mathcal{A}V(\bm r)=-\infty$ for all~$\xi\in(0,1)$.
It is worth noting that this latter result relies on the fact 
that 
$d_{11}(\bm r)=O(\vert r_1\vert^{1+\xi})$ as~$\Vert\bm r\Vert\to\infty$
with~$1+\xi>1$.

Property~2 is a consequence of the continuity of~$\mathcal{A}V$ and of property~1.

Finally, the transition semigroup satisfies
\[
T_t f(\bm r)=f(\bm r)+\int_0^t T_s\mathcal{A}f(\bm r)ds
\]
for any~$f\in\mathscr{C}^2_b(\Omega)$. The same property holds true for the
function~$V$,
as can be shown using an approximation procedure similar to the one described
in ref.~\cite{P01}, pp. 167--168. The details are given in appendix~\ref{app:B}.
\qed
\end{proof}
We now make use of the properties of the Lyapunov function to obtain the following result.
\begin{proposition}
There exists an invariant measure~$\mu$ for the stochastic process~$\bm R$, i.e.
\begin{equation}
\label{eq:invariant}
\int_\Omega\big(T_t f\big)(\bm r)\,\mu(d\bm r)=
\int_\Omega f(\bm r) \,\mu(d\bm r)
\end{equation}
for all~$f\in\mathscr{C}_b(\Omega)$ and for all~$t>0$.
\end{proposition}
\begin{proof}
Given~$\bm r_0\in\Omega$, we have
\begin{multline*}
\dfrac{1}{t}\int_0^t T_s(m-\mathcal{A}V)(\bm r_0)ds=
m-\dfrac{1}{t}\int_0^t T_s\mathcal{A}V(\bm r_0)ds\\
=m+\dfrac{V(\bm r_0)-T_t V(\bm r_0)}{t}
\le m+\dfrac{V(\bm r_0)}{t}.
\end{multline*}
Hence
\begin{equation}
\label{eq:sup}
\sup_{t\ge 1}\left[\dfrac{1}{t}\int^t_0T_s(m-\mathcal{A}V)(\bm r_0)ds\right]
<\infty.
\end{equation}
We now introduce the family
of ``average'' measures~$(\mu_t)_{t\ge 1}$ on~$\Omega$ defined,
for any~$f\in\mathscr{C}_b(\Omega)$, as
\[
\int_\Omega f(\bm r)\, \mu_t(d\bm r)=\dfrac{1}{t}\int_0^t 
T_s f(\bm r_0)ds.
\]
We show that~$(\mu_t)_{t\ge 1}$ is uniformly tight. For
a given~$N>0$ we define
\[
E_N=\{\bm r\in\Omega : m-\mathcal{A}V(\bm r)\le N\}.
\]
The set~$E_N$ is compact: it is 
closed since it is the preimage of a closed subset 
of~$\Omega$, and must be bounded since~$\mathcal{A}V(\bm r)\to -\infty$
as~$\Vert\bm r\Vert\to\infty$. 
As a consequence of Markov's inequality, 
the measure of the complement of~$E_N$ satisfies:
\[
\mu_t(E_N^c)\leq \dfrac{1}{N} \int_\Omega
[m-\mathcal{A}V(\bm r)]\mu_t(d\bm r)=
\dfrac{1}{Nt}\int_0^t T_s(m-\mathcal{A}V)(\bm r_0)ds.
\]
Using eq.~\eqref{eq:sup} we conclude that for all~$\epsilon>0$
there exists~$E_N\subset\Omega$ with
\[
N=\dfrac{1}{\epsilon}\,
\sup_{t\ge 1}\left[\dfrac{1}{t}\int^t_0
T_s(m-\mathcal{A}V)(\bm r_0)ds\right]
\]
such that~$\mu_t(E_N^c)\le\epsilon$ for all~$t\ge 1$. 
The family~$(\mu_t)_{t\ge 1}$ is therefore uniformly tight.
Then, there exists a measure~$\mu$ and a sequence~$(t_p)_{p\ge 0}$ with
$\lim_{p\to\infty}t_p=\infty$ such 
that~$\mu_{t_p}$ converges weakly to~$\mu$
as~$p\to\infty$. This means that
$\int_\Omega f\,d\mu_{t_p}\to\int_\Omega f\,d\mu$
as~$p\to\infty$ for all~$f\in\mathscr{C}_b(\Omega)$
(e.g., ref.~\cite{D89}, theorem~11.5.4, p.~404).

We now show that~$\mu$ is invariant. For any~$f\in\mathscr{C}_b(\Omega)$, 
for~$t>0$, and for 
all~$p$ such that~$t_p\ge t$, we have:
\begin{multline*}
\left\vert\int_\Omega f(\bm r)\mu_{t_p}(d\bm r)-
\int_\Omega T_t f(\bm r)\mu_{t_p}(d\bm r)\right\vert
\\
=\left\vert\dfrac{1}{t_p}\int_0^{t_p} 
T_s f(\bm r_0)\, ds
-\dfrac{1}{t_p}\int_0^{t_p} 
T_s T_t f(\bm r_0)\, ds\right\vert
\\
=\left\vert\dfrac{1}{t_p}\int_0^{t_p} T_s f(\bm r_0)\, ds
-\dfrac{1}{t_p}\int_t^{t+t_p} T_s f(\bm r_0)\, ds\right\vert
\\
=\left\vert\dfrac{1}{t_p}\int_0^{t}T_s f(\bm r_0)\, ds
-\dfrac{1}{t_p}\int_{t_p}^{t+t_p}T_s f(\bm r_0)\, ds\right\vert
\le \dfrac{2t\Vert f\Vert_\infty}{t_p}.
\end{multline*}
Hence, for all~$t>0$,
\[
\int_\Omega f(\bm r) \,\mu_{t_p}(d\bm r)-
\int_\Omega T_t f(\bm r)\,
\mu_{t_p}(d\bm r)\to 0 \qquad \text{as $p\to \infty$.}
\]
The semigroup~$(T_t)_{t\ge 0}$ satisfies
$T_t(\mathscr{C}_b(\Omega))\subset\mathscr{C}_b(\Omega)$
since $(T)_{t\ge 0}$ has the strong Feller property.
By using the weak convergence of~$\mu_{t_p}$ to~$\mu$, we can
thus conclude that~\eqref{eq:invariant} holds
for all~$f\in\mathscr{C}_b(\Omega)$ and for all~$t>0$.
The measure~$\mu$ is therefore invariant for~$\bm R$.
\qed
\end{proof}
To show that the invariant measure is actually unique,
we need the following result stating that~$\bm R$ has no
closed invariant set different from the whole space.
\begin{lemma}
\label{lemma:irreducible}
The
semigroup~$(T_t)_{t\ge 0}$ is irreducible, i.e. the
transition probabilities of~$\bm R$, $P(0,\bm r;t,U)$, 
are strictly positive for
all~$t>0$, for all~$\bm r\in\Omega$, and for all non-empty open sets~$U\subseteq\Omega$.
\end{lemma}
\begin{proof}
For all~$\bm r\in\mathbb{R}^2$ the 
linear transformation associated with~$\tilde{\sigma}(\bm r)$ is invertible, and therefore 
maps~$\mathbb{R}^2$ into itself. 
Hence, the semigroup~$\big(\widetilde{T}_t\big)_{t\geq 0}$ is irreducible (ref.~\cite{S89}, theorem~24, p. 66). 

The transition probabilities of~$\bm R$ are connected to those of~$\widetilde{\bm R}$
as follows: $P(0,\bm r;t,U)=\widetilde{P}(0,\bm r^*;t,U^*)$ 
where~$U^*=\mathfrak{p}^{-1}(U)$ and~$\bm r^*$ is any point in~$\mathfrak{p}^{-1}(\{\bm r\})$.
Therefore, $(T_t)_{t\geq 0}$ is irreducible. \qed
\end{proof}
We can now state the main result regarding the invariant measure of~$\bm R$.
\begin{theorem}
\label{th:ergodic}
There exists a unique invariant measure~$\mu$ for the 
stochastic process~$\bm R$.
The measure~$\mu$ is ergodic and equivalent to any transition 
probability $P(0,\bm r;t,U)$ with~$\bm r\in\Omega$, $t>0$, 
and~$U\subseteq\Omega$ measurable. 
Moreover, $\mu$ is absolutely continuous with respect to
the Lebesgue measure, and is therefore non degenerate (i.e., broad in~$\bm r$).
\end{theorem}
\begin{proof}
We have already proved that~$\mu$ is invariant.
Its uniqueness,
ergodicity, and equivalence to any transition probability follow
from the fact that
the transition semigroup associated with~$\bm R$ has the strong Feller property
and is irreducible~\cite{D48,K60} (see also ref.~\cite{DPZ96},
chapter~4).

To prove the absolute continuity of~$\mu$ with respect to the
Lebesgue measure, we introduce the family of transition probabilities
\[
Q_{\lambda}(\bm r,U)=\lambda\int_0^\infty e^{-\lambda s}
P(0,\bm r;s,U) \, ds
\]
with~$\bm r\in\Omega$ and~$U\subseteq\Omega$ measurable, as well as the 
associate transition semigroup
\[
\mathcal{T}_\lambda f(\bm r)=\int_\Omega f(\bm y)Q_\lambda(\bm r,d\bm y).
\]
Likewise, we define an analogous family~$\widetilde{Q}_{\lambda}(\bm r,U)$
for the process~$\widetilde{\bm R}$.
The measure~$\mu$ is invariant also for~$(\mathcal{T}_\lambda)_{\lambda\geq 0}$:
\[
\int_{\Omega}\mu(d\bm y)\mathcal{T}_\lambda f(\bm y) =
\lambda\int_0^\infty ds\, e^{-\lambda s}\int_\Omega \mu(d\bm y)
T_s f(\bm y)=\int_\Omega \mu(d\bm y)f(\bm y)
\]
for any~$f\in\mathscr{B}_b(\Omega)$, and hence
\begin{equation}
\label{eq:abs-cont}
\mu(U)=\int_\Omega \mu(d\bm y)Q_\lambda(\bm y,U)
\end{equation}
for any measurable set~$U\subseteq\Omega$.

For all~$\bm r\in\mathbb{R}^2$, the 
measure~$\widetilde{Q}_{\lambda}(\bm r,\cdot)$ is absolutely continuous
with respect to the Lebesgue measure (see ref.~\cite{S89}, theorem~10, p.~24).
It follows that~$Q_\lambda(\bm r,\cdot)$ has the same property for 
all~$\bm r\in\Omega$ since
$Q_\lambda(\bm r,U)=\widetilde{Q}_\lambda
(\bm r^*,\mathfrak{p}^{-1}(U))$ 
for a given~$\bm r^*\in\mathfrak{p}^{-1}\{\bm r\}$. 
From eq.~\eqref{eq:abs-cont}, $\mu$
is therefore absolutely continuous with respect to the Lebesgue measure. \qed
\end{proof}
As a consequence of lemma~\ref{lemma:irreducible} 
and theorem~\ref{th:ergodic}, the transition probability of~$\bm R$
has a positive density with respect to the Lebesgue measure:
$P(0,\bm r;t,d\bm\rho)=p(0,\bm r;t,\bm\rho)d\bm\rho$.
The probability density function is the (possibly weak) solution of:
\begin{equation}
\label{eq:pdf}
\partial_t p=\mathcal{M} p,
\end{equation}
where, for~$f\in \mathscr{C}^2(\Omega)$,
\[
\mathcal{M} f(\bm\rho)=\sum_{1\le\alpha,\beta\le 2}
\partial_{\rho_\alpha}\partial_{\rho_\beta}d^\kappa_{\alpha\beta}(\bm\rho)f(\bm\rho). 
\]

\section{Conclusions}
\label{sec:conclusions}
We have studied the dynamics of fluid particles in a compressible 
turbulent velocity field  on a cylinder. The model that we have introduced is a generalization
of the isotropic Kraichnan ensemble. Although
the parameters of the velocity
have been set in such a way as
to produce explosive separation  of the fluid particles in the isotropic limit
($L\to\infty$),
on the cylinder the probability distribution of the separation
tends to an invariant measure. This behavior is a result of the 
compressibility effects generated at large scales by the compactification of the ``radial'' dimension.

The diffusivity~$\kappa$ has been taken strictly positive to
guarantee the existence of solutions to eq.~\eqref{eq:weak}.
The addition of Brownian motion to Lagrangian trajectories
influences the dynamics of fluid particles at small separations.
Therefore, the presence of a nonzero diffusivity
may be relevant for the non-degeneracy of the invariant measure, but should not affect
the existence of the invariant measure itself, which rather depends on the large-scale form
of the velocity field.
The limit~$\kappa\to 0$ 
may be tackled by means of Wiener chaos decomposition methods~\cite{LJR98,LJR02,LJR04,LJR05}. 
We conjecture that our results remain valid in that limit. Indeed, in 
the situation considered, the small-scale dynamics of fluid particles is the same as
in the weakly compressible phase of the isotropic Kraichnan ensemble.
In that regime,  Lagrangian trajectories separate in time even for 
vanishing~$\kappa$
owing to the poor spatial regularity of the velocity~\cite{GV00}. 
Thus, the invariant measure should remain non-degenerate as~$\kappa\to 0$. 

The Reynolds number is infinite in our study since the viscosity 
of the fluid, $\nu$, has been set to zero from the beginning.
For the same reason the Prandtl number~$\mathit{Pr}=\nu/\kappa$ is equal to zero.
The viscosity can be taken into account by multiplying the spectral 
tensor~\eqref{eq:spectral-tensor} by the
factor~$e^{-\eta^2\Vert\bm k\Vert^2}$, where~$\eta\propto\nu^{3/4}$ 
plays the role of the viscous length of the flow~\cite{FGV01}.
This modification has a small-scale regularizing effect on 
the velocity field, which for any~$\eta>0$
is locally Lipschitz continuous.
Obviously, a positive~$\eta$ does not alter the proofs of the results shown in the paper.

The order of the limits~$\kappa\to 0$ and~$\eta\to 0$, however,
deserves a detailed discussion.
Taking the limit~$\kappa\to 0$ before~$\eta\to 0$ is equivalent to letting~$\mathit{Pr}$ tend to infinity.
The opposite order corresponds to the limit~$\mathit{Pr}\to 0$.
As first observed in ref.~\cite{EVE00}, when these limits are considered
the range of weak compressibility splits into two ranges:
what is now called the range of weak compressibility in the strict sense, 
$0\le\wp<(d-2+\xi)/(2\xi)$, and the range of \textit{intermediate}
compressibility, $(d-2+\xi)/(2\xi)\le\wp<d/\xi^2$.
In the former range, 
the order of the limits~$\kappa\to 0$ and~$\eta\to 0$ is not relevant for the Lagrangian dynamics~\cite{GH04}.
At small scales, fluid particles disperse irrespective of the order of the limits, and therefore
we expect the invariant measure of the separation  
to be non-degenerate.
By contrast, the order matters in the latter range~\cite{GH04}.
For intermediate values of the compressibility,
if the viscous regularization is removed before the
diffusivity ($\mathit{Pr}\to 0$),
the small-scale Lagrangian dynamics is once more characterized by the explosive separation of the trajectories.
If~$\kappa$ goes to zero before~$\eta$ ($\mathit{Pr}\to\infty$), the trajectories
coalesce also at small scales,
and the invariant measure of the separation should degenerate into a Dirac delta function. 

In summary, we believe that the present study captures the behavior of the
Lagrangian trajectories on cylindrical manifolds for all~$Pr$
except for the limit~$\mathit{Pr}\to\infty$ in 
the  intermediate-compressibility regime.
These results are, moreover, relevant to
turbulent transport of passive scalar fields
in virtue of the relation subsisting
between the scalar correlations
and the dynamics of fluid particles~\cite{GV00,FGV01}.

We conclude by noting that when both the dimensions of the plane
are compactified one obtains the Kraichnan flow on a two-dimensional periodic
square studied in ref.~\cite{CDG07}. The velocity field considered there
was however smooth in space.

\begin{acknowledgements}
The authors are grateful to F.~Flandoli and Y.~Le~Jan
for fruitful discussions.
\end{acknowledgements}

\appendix

\section{Covariance of velocity differences}
\label{app:A}

For~$d=2$, $d'=1$, $\xi\in(0,1)$, $\xi'\in (1,2)$,
the spatial covariance of the velocity field takes the form
$$
D_{\alpha\beta}(\bm r)=\frac{1}{4\pi^2 L}
\sum_{j=-\infty}^{\infty} e^{i\frac{j}{L}r_2}\int_{\mathbb{R}} dk_1\,
\frac{e^{ik_1 r_1}A_{\alpha\beta}((k_1,\frac{j}{L});\wp)}%
{\left(k_1^2+\frac{j^2}{L^2}+\frac{1}{\ell^{2}}\right)^{\frac{2+\xi}{2}}}
$$
with~$\bm r\in\Omega$.
We first establish the convergence of the above series.
It is convenient to denote
the integrals  by~$D_{\alpha\beta}^{(j)}(r_1)$
and thus rewrite the covariance as follows:
\begin{equation}
\label{eq:cov-series}
D_{\alpha\beta}({\bm r})= \frac{1}{4\pi^2L} \sum_{j=-\infty}^{\infty} D_{\alpha\beta}^{(j)}(r_1) e^{i\frac{j}{L}r_2}=
\frac{ D_{\alpha\beta}^{(0)}(r_1)}{4\pi^2L}+ \frac{1}{2\pi^2L}
\sum_{j=1}^{\infty} D_{\alpha\beta}^{(j)}(r_1) \cos\left(\frac{jr_2}{L}\right).
\end{equation}
Using the inequality
\[
\vert A_{\alpha\beta}(\bm k);\wp)\vert
\leq 1-\wp+\vert 2\wp-1\vert \qquad 
\forall\,\alpha,\beta=1,2 \quad \text{and} \quad
\forall\,\bm k\in\mathbb{R}\times \frac{1}{L}\mathbb{Z},
\]
we obtain that the coefficients of the series satisfy for all~$r_1$
\[
\vert  D_{\alpha\beta}^{(j)}(r_1)\vert \leq 
(1-\wp+\vert 2\wp-1\vert)M_j
\]
with (e.g., ref.~\cite{E53}, formula~I~1.5(2)) 
\[
M_j=\int_{\mathbb{R}} dk_1\,
\left(k_1^2+\frac{j^2}{L^2}+\frac{1}{\ell^2}\right)^{-\frac{2+\xi}{2}}=
\frac{\sqrt{\pi}\Gamma\left(\frac{1+\xi}{2}\right)}%
{\Gamma\left(1+\frac{\xi}{2}\right)}
\left(\frac{j^2}{L^2}+\frac{1}{\ell^2}\right)^{-\frac{1+\xi}{2}}.
\]
The series~$\sum_{j=1}^\infty M_j$ converges for all~$\xi\in(0,1)$
and~$\ell>0$ 
as well as in the limit~$\ell\to\infty$,
as it can be checked 
by means of the integral test. Then, the Weierstrass criterion guarantees that
the series in the right-hand-side of eq.~\eqref{eq:cov-series}
converge uniformly and absolutely on~$\Omega$. 
The uniform convergence will allow us
to compute~$\lim_{\ell\to\infty}d_{\alpha\beta}(\bm r)$ by exchanging limit and
summation.

The basic analytical ingredient to derive eqs.~\eqref{eq:d11}--\eqref{eq:d12}
is 
\begin{equation}
\label{eq:bessel}
\int_{-\infty}^{\infty} \frac{e^{ik_1r_1}}{(k_1^2+z^2)^{\nu+1/2}} dk_1 
=\frac{2^{1-\nu} \pi^{1/2} K_\nu (\vert zr_1\vert)\;\vert r_1\vert^{\nu} } {\Gamma\left(\nu+\frac{1}{2}\right)
\vert z\vert^\nu}
\end{equation}
with~$Re(\nu)>-1/2$ and $\vert\arg z\vert<\pi/2$ (e.g., ref.~\cite{E53}, 
formula II~7.12(27)).

We first compute the correlation of the axial component of the velocity; the correlation of the
other components may be easily derived from~$D_{11}(\bm r)$.

In the limit $\ell\to\infty$, we have
\begin{multline*}
\lim_{\ell\to\infty}D_{11}^{(j\ne 0)}(r_1)= \int_{-\infty}^{\infty} 
dk_1\, \frac{e^{ik_1 r_1}\left[(1-\wp)\frac{j^2}{L^2}+\wp k_1^2\right]}{\left(k_1^2+\frac{j^2}{L^2}
\right)^{2+\frac{\xi}{2}}}
\\
=(1-\wp)\,\frac{j^2}{L^2} \int_{-\infty}^{\infty} 
dk_1\, \frac{e^{ik_1r_1}}{\left(k_1^2+ \frac{j^2}{L^2}
\right)^{2+\frac{\xi}{2}}} +
\wp \int_{-\infty}^{\infty} 
dk_1\, \frac{k_1^2 e^{ik_1r_1}}{\left(k_1^2+\frac{j^2}{L^2}
\right)^{2+\frac{\xi}{2}}} 
\\[0.2cm]
= (1-2\wp)\,\frac{j^2}{L^2} \int_{-\infty}^{\infty} 
dk_1\, \frac{e^{ik_1r_1}}{\left(k_1^2+\frac{j^2}{L^2}
\right)^{2+\frac{\xi}{2}}} +
\wp \int_{-\infty}^{\infty} 
dk_1\, \frac{e^{ik_1r_1}}{\left(k_1^2+ \frac{j^2}{L^2}\right)^{1+\frac{\xi}{2}}} 
\\[0.2cm]
=(1-2\wp)\,\left|\frac{j}{L}\right|^{\frac{1-\xi}{2}} 
\frac{\pi^{1/2}\,\vert r_1\vert^{(3+\xi)/2} K_{\frac{3+\xi}{2}} \left(\left\vert \frac{j}{L}\,r_1\right\vert\right) }%
{2^{(1+\xi)/2}\,\Gamma\left(2+\frac{\xi}{2}\right)} 
\\
+
\wp\,\left|\frac{j}{L}\right|^{-\frac{1+\xi}{2}} 
\frac{2^{(1-\xi)/2}\pi^{1/2}\, \vert r_1\vert^{(1+\xi)/2}
K_{\frac{1+\xi}{2}} \left(\left\vert\frac{j}{L}\,r_1\right\vert\right)}{\Gamma\left(1+\frac{\xi}{2}\right)}. 
\end{multline*}
To compute~$D_{11}^{(j)}(0)$ we can use the asymptotic expansion 
of~$K_\nu(z)$ for~$z\to 0$
\begin{align}
\label{eq:expansion_small_1}
K_\nu(x)&\sim \frac{\Gamma(\nu)}{2}\left(\frac{x}{2}\right)^{-\nu}
+\frac{\Gamma(-\nu)}{2} \left(\frac{x}{2}\right)^{\nu} + O(x^{2-\nu})
& (\nu<1)&
\\
\label{eq:expansion_small_2}
K_{\nu}(x)&\sim \frac{\Gamma(\nu)}{2}\left(\frac{x}{2}\right)^{-\nu}-\frac{\Gamma(\nu)}{2(\nu-1)}\,
\left(\frac{x}{2}\right)^{2-\nu}+O(x^\nu)
& (1<\nu<2)&.
\end{align}
Hence we obtain
\begin{multline*}
\lim_{\ell\to\infty}D_{11}^{(j\ne 0)}(0)= 
\wp\,\left|\frac{j}{L}\right|^{-1-\xi} 
\frac{\pi^{1/2} \Gamma\left(\frac{1+\xi}{2}\right)}{\Gamma\left(1+\frac{\xi}{2}\right)} 
+(1-2\wp)\,\left|\frac{j}{L}\right|^{-1-\xi} 
\frac{\pi^{1/2} \Gamma\left(\frac{3+\xi}{2}\right)}{\Gamma\left(2+\frac{\xi}{2}\right)} 
\\
= \left|\frac{j}{L}\right|^{-1-\xi} \frac{\pi^{1/2} (1+\xi-\wp\xi)
\Gamma\left(\frac{1+\xi}{2}\right)}{2 \Gamma\left(2+\frac{\xi}{2}\right)}.
\end{multline*}
For $j=0$ and~$\ell<\infty$, we have
$$
D_{11}^{(0)}(r_1)=\wp\int_{-\infty}^{\infty}dk_1\,\frac{e^{ik_1r_1}}%
{(k_1^2+\ell^{-2})^{\frac{2+\xi}{2}}}=
\wp\,\frac{2^{\frac{1-\xi}{2}}\pi^{1/2}}{\Gamma\left(1+\frac{\xi}{2}\right)} \left\vert
\ell r_1\right\vert^{\frac{1+\xi}{2}}
K_{\frac{1+\xi}{2}}\Big(\Big\vert\frac{r_1}{\ell}\Big\vert\Big).
$$
The asymptotic expansion~\eqref{eq:expansion_small_1} 
shows that $D_{11}^{(0)}(0)$
diverges like $\ell^{1+\xi}$ as $\ell\to\infty$. By using expansion~\eqref{eq:expansion_small_1},
it is nonetheless possible to show that
$$
\lim_{\ell\to\infty}[D_{11}^{(0)}(0)-D_{11}^{(0)}(r_1)]=\wp\,\frac{\pi^{1/2}\left\vert\Gamma\left(-\frac{1+\xi}{2}\right)\right\vert}%
{2^{1+\xi}\Gamma\left(1+\frac{\xi}{2}\right)}\,\vert r_1\vert^{1+\xi}
$$ 
(note that $\Gamma(-(1+\xi)/2)<0$ for $0<\xi<1$). Hence, the covariance
of the axial component of the velocity difference, $d_{11}(\bm r)$,
has a finite limit as~$\ell\to\infty$.

For the other components we have:
$$
D_{22}^{(j)}(r_1)= \int_{-\infty}^{\infty} 
dk_1\, \frac{e^{ik_1r_1}\left[(1-\wp)k_1^2+\wp\,\frac{j^2}{L^2}\right]}{\left(k_1^2+ \frac{j^2}{L^2}\right)
\left(k_1^2+ \frac{j^2}{L^2}+\frac{1}{\ell^2}\right)^{\frac{2+\xi}{2}}}
$$
and
$$
D_{12}^{(j)}(r_1)= D_{21}^{(j)}(r_1)=(2\wp-1)\,\frac{j}{L}
\int_{-\infty}^{\infty} 
dk_1\, \frac{k_1 e^{ik_1r_1}}{\left(k_1^2+ \frac{j^2}{L^2}\right)
\left(k_1^2+ \frac{j^2}{L^2}+\frac{1}{\ell^2}\right)^{\frac{2+\xi}{2}}}\, .
$$
Therefore, $D_{22}^{(j)}(r_1)$ can be derived from $D_{11}^{(j)}(r_1)$ by replacing $\wp$ with 
$1-\wp$:
\[
\begin{array}{l}\displaystyle
\begin{split}
\lim_{\ell\to\infty}D_{22}^{(j\ne 0)}(r_1)&=
(2\wp-1)\, \left|\frac{j}{L}\right|^{\frac{1-\xi}{2}} 
\frac{\pi^{1/2} K_{\frac{3+\xi}{2}} \left(\big\vert \frac{j}{L}r_1\big\vert\right)\; 
\vert r_1\vert^{(3+\xi)/2} } {2^{(1+\xi)/2}\Gamma\left(2+\frac{\xi}{2}\right)}
\\
&+(1-\wp)\,\left|\frac{j}{L}\right|^{-\frac{1+\xi}{2}} 
\frac{2^{(1-\xi)/2}\pi^{1/2} K_{\frac{1+\xi}{2}} \left(\big\vert \frac{j}{L}r_1\big\vert\right)\; 
\vert r_1\vert^{(1+\xi)/2} } {\Gamma\left(1+\frac{\xi}{2}\right)}, 
\end{split}
\\[1.3cm]
\displaystyle\lim_{\ell\to\infty}
D_{22}^{(j\ne 0)}(0) 
=\left\vert\frac{j}{L}\right\vert^{-1-\xi} \frac{\sqrt{\pi}\,(1+\wp\xi)
\Gamma\left(\frac{1+\xi}{2}\right)}{2 \Gamma\left(2+\frac{\xi}{2}\right)},
\\[0.8cm]
\displaystyle\lim_{\ell\to\infty}[D_{22}^{(0)}(0)-D_{22}^{(0)}(r_1)]=(1-\wp)\,\frac{\pi^{1/2}
\left\vert\Gamma\left(-\frac{1+\xi}{2}\right)\right\vert}%
{2^{1+\xi}\Gamma\left(1+\frac{\xi}{2}\right)}\,\vert r_1\vert^{1+\xi}.
\end{array}
\] 
The mixed correlation can be obtained, for $j\neq 0$, by differentiating formula~\eqref{eq:bessel} with respect to~$r_1$
and by using $\frac{d}{dx} [x^\nu K_\nu(x)]=-x^\nu K_{\nu-1}(x)$
(e.g., ref~\cite{E53}, formula~II~7.11(21)):
$$
\lim_{\ell\to\infty} D_{12}^{(j\ne 0)}(r_1)= i (2\wp-1)\frac{j}{L}\,\left\vert\frac{j}{L}\right\vert^{-\frac{1+\xi}{2}} 
\frac{ \sqrt{\pi} \, \vert r_1\vert^{(3+\xi)/2}}{2^{\frac{1+\xi}{2}}\Gamma\left(2+\frac{\xi}{2}\right) }\,
K_{\frac{1+\xi}{2}}\left(\vert\textstyle\frac{j}{L}r_1\vert\right) \text{sgn}(r_1).
$$
Hence
$$
\lim_{\ell\to\infty} D_{12}^{(j\ne 0)}(0)=0.
$$
For $j=0$ we have
$$
D_{12}^{(0)}(r_1)=0 \qquad \forall\, r_1\in\mathbb{R}.
$$
Finally, eqs.~\eqref{eq:d11}--\eqref{eq:d12} may be derived by
recalling that 
$$
d_{\alpha\beta}({\bm r})=\frac{1}{4\pi^2 L} \sum_{j=-\infty}^{\infty}
\left[ D^{(j)}_{\alpha\beta}(0) -D^{(j)}_{\alpha\beta}(r_1)e^{i \frac{j}{L} r_2}
\right]
$$
and using the uniform convergence of the series in the right-hand-side of
eq.~\eqref{eq:cov-series}.

\section{Proof of Lemma~\ref{lemma-V}}
\label{app:B}

To prove property~3 of lemma~\ref{lemma-V}, we first observe that
$$
\mathbb{E}\left(\Vert \bm R(t)\Vert^2\right)=
\Vert\bm R(0)\Vert^2 + 2\,\mathbb{E}\left(\int_0^t \Vert \sigma
(\bm R(s))\Vert^2 ds\right) \leq
C_2\left(1+\int_0^t \mathbb{E}\left(\Vert \bm R(s)\Vert^2\right)ds\right),
$$
where~$C_2>0$ and ~$\Vert\sigma\Vert:=
\big[\mathrm{Tr}\big(\sigma\sigma^{\mathrm{T}}\big)\big]^{1/2}$.
The above inequality is a consequence of~\eqref{eq:d-asympt}.
Gronwall's inequality then yields
\begin{equation}
\label{eq:Gronwall}
\mathbb{E}\left(\Vert \bm R(t)\Vert^2\right) \leq C_2 e^{C_2t}
\end{equation}
for all~$t>0$. 

Following ref.~\cite{P01}, we consider, for 
all~$\gamma>0$, the function $\varphi_\gamma:\mathbb{R}_+\to\mathbb{R}_+$ with
\[
\varphi_\gamma(z)=
\begin{cases}
z & 0\leq z\leq\gamma 
\\
\varphi_\gamma(\gamma+1) & \gamma+1\leq z
\end{cases}
\]
and~$\varphi_\gamma\in\mathscr{C}^2(\mathbb{R}_+)$ and monotonically 
non-decreasing.
Moreover, we define~$V_\gamma:=\varphi_\gamma\circ V$.
Applying~$\mathcal{A}$ to~$V_\gamma$ 
and taking into account~\eqref{eq:d11-asymptotic}
yield
\begin{equation}
\label{eq:AV-gamma}
\vert\mathcal{A}V_\gamma(\bm r)\vert=\sigma_{11}^2(\bm r)
\left\vert
\varphi_\gamma'(V(\bm r))\,\dfrac{\partial^2 V}{\partial r_1^2}+
\varphi_\gamma''(V(\bm r))\left(\dfrac{\partial V}{\partial r_1}\right)^2
\right\vert = O\left(\vert r_1\vert^{3+\xi-4h}\right)
\end{equation}
with~$3+\xi-4h<2$ as~$\Vert\bm r\Vert\to\infty$.

Since~$V_\gamma\in\mathscr{C}^2_b(\Omega)$, we have for all~$\gamma>0$
\[
T_t V_\gamma(\bm r)=V_\gamma(\bm r)+\int_0^t T_s(\mathcal{A}V_\gamma)(\bm r)ds.
\]
We now show that each term of the above equation tends as~$\gamma\to\infty$
to the corresponding term in property~3 of lemma~\ref{lemma-V}.

Obviously, $\lim_{\gamma\to\infty}V_\gamma(\bm r)=V(\bm r)$ for 
all~$\bm r\in\Omega$. Likewise, $V_\gamma(\bm R(t))\nearrow V(\bm R(t))$
almost everywhere as~$\gamma\to\infty$. Therefore, by the monotone
convergence theorem 
$\lim_{\gamma\to\infty}T_t V_\gamma(\bm r)=T_t V(\bm r)$.
Finally, $\lim_{\gamma\to\infty}\mathcal{A}V_\gamma(\bm R(t))=\mathcal{A}V(\bm R(t))$
and, from eqs.~\eqref{eq:AV-gamma} and~\eqref{eq:Gronwall}, 
$\vert AV_\gamma(\bm R(t))\vert \leq C_3(1+\Vert\bm R(t)\Vert^2)$ with~$C_3>0$
and $\mathbb{E}\left(\Vert \bm R(t)\Vert^2\right)<\infty$.
Then, it follows from the bounded convergence theorem that
$$
\lim_{\gamma\to\infty}\int_0^t T_s(\mathcal{A}V_\gamma)(\bm r)ds=
\int_0^t T_s(\mathcal{A}V)(\bm r)ds.
$$
This concludes the proof.



\end{document}